\documentclass[preprint,showpacs]{revtex4-1}

\usepackage{amsmath}
\usepackage{amssymb}
\usepackage{graphicx}

\newcommand{\qket}[1]{\left| #1 \right>}

\newcommand{\qexp}[3]{\left< #1 \left| \vphantom{#1 #3} #2 \right| #3 \right>}

\newcommand{\rd}{\ensuremath{{\rm d}}}
\newcommand{\avg}[1]{\ensuremath{\left\langle{#1}\right\rangle}}

\newcommand{\kBT}{\ensuremath{k_{\text{B}}T}}

\begin{document}
\title[Activity and Kinetics of 1D Ising Model]{%
Activity and  Steady-State Kinetics of\\ One-Dimensional Kinetic Ising Model}
\author{Kuo Kan \surname{Liang}}
\affiliation{$^1$ Research Center for Applied Sciences, Academia Sinica, 128 Academia Road, Section 2, Nankang, Taipei 115-29, Taiwan}
\affiliation{$^2$ Department of Biochemical Science and Technology, National Taiwan University, No. 1, Sec. 4, Roosevelt Road, Taipei, 10617 Taiwan}
\email{kkliang@sinica.edu.tw}

\begin{abstract}
In this work we focused on the kinetics of a one-dimensional Ising system (1DIS) 
with constant nearest-neighbor interaction (NNI).
The exact solution of both thermodynamics and kinetics of this system under 
quasi-chemical approximation (QCA) had been shown
in the literature, and the equilibrium solution was exact.
In this work, it was discussed why QCA applied the best in the case of 1DIS with constant NNI.
Furthermore, extension had been made to discuss that due to this special reason, 
perhaps the kinetics of the system under QCA is the correct steady-state kinetics.
Inspired by this observation, the activity and activity coefficients of the system was studied closely
to re-examine the form of the equation of motion under QCA.
A novel concept---the instantaneous activities and the corresponding 
instantaneous activity coefficients---was introduced,
and in terms of these quantities
the kinetics seemed to be much simpler and physically more meaningful.
The chevron plot of this system was also discussed and new way of looking at the
\emph{rollover} of chevron plots was presented.
\end{abstract}
\maketitle

\section{Introduction}
The Ising model is a macroscopic system of interacting constituents with the 
fewest possible system variables.
It provides the insight to the essential characters of phase transition, 
and more generally the cooperative phenomena.
In the past two decades, Ising-like models were also widely applied in protein sciences.
According to the level of specificity, there were two major classes of models.
The one-dimensional model of Zwanzig\cite{pnas92:9801} is highly abstract one, and the 
Wako-Sait\^o-Mu\~noz-Eaton (WSME) model\cite{pnas95:5872,pnas96:11311},
also one-dimensional, was very concrete.
The present author and coworkers had employed a model with complexity in between that
of Zwanzig and WSME models\cite{pccp5:5300,jccs50:335,jccs51:1161,cpl399:440,jac40:s195,%
bj94:4828,jbp38:543}.
The advantage of such models is that it emphasizes on some properties
common to all different realistic systems, while it does that with clear enough details.

In this work, the system discussed is a one-dimensional Ising model
with constant nearest neighbor interaction (NNI).
The components of this model are related to the \emph{discrete parameters}
introduced by Zwanzig\cite{pnas92:9801} and are called the \emph{structural units} (SU).
In analogy with the magnetic systems, 
in which an external magnetic field can create energy difference between 
the spin up and down state,
the external conditions such as temperature, pH, and denaturant concentration can induce 
a free energy difference between the native (up) and denatured (down) states of an SU.
In this work, this energy difference is called the
\emph{local free energy difference}.
Interactions of a specific SU with its neighbors also result in an energy difference between
the native and denatured states, 
and this energy difference obviously depends not only on the state of this SU, 
but also on the states of its neighbors.
This interaction energy, together with the local free energy difference 
under the external conditions, determines the overall energy difference 
between the native and denatured states of this SU.
System-wise, the total energy depends on the instantaneous external conditions,
and on the instantaneous structural configuration of the system through a parameter 
describing the strength of the NNI.
The exact equilibrium solution of the system, specifically, the partition function of it,
can be obtained by standard method of transfer matrix\cite{pr60:252},
and it will reduce to a very simple form by assuming that the system size goes to infinity.

The kinetics of such systems had been widely studied, and perhaps most extensively with
phenomenological approaches.
Starting with the master equation approaches,
assumption can be made on the relations between the transition rates and the
characteristic time scale of the processes\cite{jmp4:294,pr149:301,jsp29:375},
or the derivation of the form of the rates from microscopic theory can be worked out%
\cite{jcp63:5116,pa137:439}.
Since these results are not the original contribution by the present author, 
detailed derivations of the theories were avoided whenever possible.
The emphasis here are the extensions and re-formulations that reveal insights
that were not noticed before.
To that end, however, those known results that are used herein should be highlighted,
and the notations be unified.

In the following Section, the theory is highlighted in four parts.
First, the notations are introduced.
Second, the exact partition function of the system is presented.
With that, the equilibrium properties of the system can be shown.
Third, the master equations are presented, and the Glauber's form of 
the principle of detailed balance\cite{jmp4:294}
is used to write down the relevant rates in the equations.
Various approximations could be applied to make the master equations more easily solvable.
In the fourth part the mean-field and the quasi-chemical approximations are introduced.
It is shown that for the constant NNI model quasi-chemical approximation results in
the correct equilibrium solution.
Therefore, in the third Section, the kinetics of the quasi-chemical model is analyzed in detail.
Specifically, the nonideality of the interacting-SU system is represented by its activities
and activity coefficients.
It is found that by using the analytic forms of the activities and coefficients to define
corresponding kinetic quantities which shall be called the ``instantaneous activities'' or
``instantaneous activity coefficients'',
the kinetics of the system can be expressed in a very elegant form.
The rate law is so simple that the chevron plot of this model can also be discussed.
In the final Section, this work is concluded with some outlooks.

\section{Theory}
In the following, the notations and major theories used throughout this text 
are described in detail.
The thermodynamics and the kinetics of the system are mostly already solved in the literature.
Therefore the details of the derivations are sometimes left out.
Although cares had been taken not to make it look like all of the contributions were
from the present author, sometimes, for the readability of the text,
distinction between known facts and new extensions is hard to make.
The author wishes to acknowledge all of the earlier contributors in this field,
and apologizes first, before the discussions proceed. 

\subsection{Notations}
The system consists of a number $M$ of identical but distinguishable components.
Each component, like a spin-$1/2$ particle, has two states.
In this work we call these \emph{local two-state} (LTS) components the
\emph{structural units} (SU).
The $j$-th SU is quantified by a normalized binary variable $\sigma_j$
that takes the value of either $+1$ or $-1$.
Topologically, this system can, by borrowing the Dirac notation, be expressed as
\begin{equation}
 \qket{\vec{\sigma}}\equiv
 \qket{\sigma_1,\ldots,\sigma_{j-1},\sigma_j,\sigma_{j+1},\ldots,\sigma_M}.
 \label{e:n001}
\end{equation}
Later the size of the system will be taken to infinity, therefore the boundary effect is
of very little concern.
In that case, circular boundary condition can be used to further simplify the model.
Consequently, a Hamiltonian-like quantity $\hat{H}$ can be defined as following:
\begin{equation}
 \hat{H}\qket{\vec{\sigma}}=
 \left[-\sum_j\sigma_j\varepsilon_j\left(\vec{\sigma}\right)\right]
 \qket{\vec{\sigma}}.
 \label{e:n002}
\end{equation}
$\hat{H}$ is said to be `Hamiltonian-like' because
actually it can represent the free energy of the system instead of just the internal energy.
Since in this simplified spin-like view of the system 
the spatial configuration is not considered at all,
the kinetics energy is implicit, and there is no expansion work.
Hence it is not necessary to distinguish between internal energy and various free energies.
The factor $\varepsilon_j\left(\vec{\sigma}\right)$ measures the energy difference
between the native and denatured states of the $j$-th SU
when the system is in a specific configuration $\vec{\sigma}$.
The circular symmetry implies that the ways to determine the energy of every SU are identical.
Moreover, a common zero of energy can be found so that, viewing from any of the SUs,
for any configuration of the remaining SUs, 
the energies corresponding to the native and denatured states
are always equal in magnitude and opposite in sign.
This justify the notation in Eq.~\eqref{e:n002}.
In the present model, indeed
\begin{equation}
\varepsilon_j\left(\vec{\sigma}\right)=\epsilon+J\left(\sigma_{j-1}+\sigma_{j+1}\right)/2
\label{e:n003}
\end{equation}
The energy term $\epsilon$ is the energy difference between the denatured
and the native states when there is zero cooperativity.
Although seemingly natural, this definition is in fact self-contradicting.
When the sign of $\epsilon$ changes, the originally native state becomes the
non-native state.
Therefore the native state should be more clearly defined as the more stable
state when $\epsilon>0$.
Thus defined, $\epsilon$ shall be called the 
\emph{local free energy difference} in later text for brevity,
because it reflects the thermodynamic properties of a local two-state system.
The other factor $J$ represents the constant coupling between neighboring SUs.
In Eq.~\eqref{e:n003}, every interaction is counted twice when the whole system is considered,
a factor of one-half is included there for compensation.
When $J$ is positive, neighboring spins of the same signs contribute a negative energy term,
meaning that this configuration is more stable than neighboring spins with opposite signs.
Therefore positive $J$ implies positive cooperativity.
Naturally, negative $J$ implies negative cooperativity.
The total energy of a configuration of the system determines the relative stability of it,
and we can sum up the contributions to obtain
\begin{equation}
 \hat{H}\qket{\vec{\sigma}}=
 \left[-\sum_j\left(\epsilon\sigma_j+J\sigma_j\sigma_{j+1}\right)\right]\qket{\vec{\sigma}}.
 \label{e:n004}
\end{equation}

\subsection{Partition function and some properties}
The partition function is the most useful and fundamental quantity of a
statistical mechanical system.
The methods to obtain the partition function and to apply it to obtain the
most important statistical properties of the present model are standard and
can be found in many textbooks and articles.
The procedures presented in this Subsection follow the text of K.~Huang\cite{ising:kersonhuang}
and the book chapter by H.~W.~Huang and W.~A.~Seitz\cite{cib:c3}.

The partition function of this system can be written as
\begin{align}
 Z&=\sum_{\left\{\vec{\sigma}\right\}}\prod_j%
e^{\left[\epsilon\sigma_j+J\sigma_j\left(\sigma_{j-1}+\sigma_{j+1}\right)/2\right]/k_{\rm B}T}
\nonumber\\
&=\sum_{\left\{\vec{\sigma}\right\}}\prod_j%
e^{\bar{\epsilon}\sigma_j+\bar{J}\sigma_j\left(\sigma_{j-1}+\sigma_{j+1}\right)/2}
\label{e:n005}
\end{align}
where for brevity we defined $\bar{\epsilon}=\epsilon/k_{\rm B}T$ and
$\bar{J}=J/k_{\rm B}T$.
Transfer matrix method can be used to calculate the analytic form of $Z$\cite{pr60:252}:
\begin{equation}
 Z=\lambda_+^M+\lambda_-^M=
\lambda_+^M\left(1+\frac{\lambda_-^M}{\lambda_+^M}\right).%
\label{e:n006}
\end{equation}
where
\begin{equation}
 \lambda_{\pm}=e^{\bar{J}}\cosh\bar{\epsilon}\pm
\sqrt{e^{2\bar{J}}\sinh^2\bar{\epsilon}+e^{-2\bar{J}}}
\label{e:n007}
\end{equation}
Since $\left|\lambda_-/\lambda_+\right|<1$ in any case, when $M$ is large,
$Z\approx\lambda_+^M$.

With the partition function, any thermo-statistical property of the system can be obtained.
The most important observable in the Ising model for a spin system is the 
`magnetization' which can be directly related to the system-average equilibrium value of $\sigma$:
\begin{equation}
\avg{\sigma}_{\text{eq}}=\frac{
\sum_{\left\{\vec{\sigma}\right\}}\sum_j\sigma_j\exp\left[%
\bar{\epsilon}\sum_k\sigma_k+%
\bar{J}\sum_k\left(\sigma_{k-1}+\sigma_{k+1}\right)\sigma_k/2\right]}{Z}
\label{e:n008}
\end{equation}
With this definition, it can be easily shown that
\begin{equation}
\avg{\sigma}_{\text{eq}}=\frac{1}{M}\,\frac{\partial\ln Z}{\partial\bar{\epsilon}}
\label{e:n009}
\end{equation}
Replacing $Z$ by $\lambda_+^M$ and use Eq.~\eqref{e:n007} it is obtained that
\begin{equation}
 \avg{\sigma}_{\text{eq}}=\frac{e^{2\bar{\epsilon}}-1}{\sqrt{\left(e^{2\bar{\epsilon}}-1\right)^2+4e^{2\bar{\epsilon}}e^{-4\bar{J}}}}.
\label{e:n010}
\end{equation}

In a spin system $\avg{\sigma}$ is the observable `magnetization'.
In other systems of collection of LTS, it is not necessarily the most relevant observable.
In contrast, the ``fraction of spins up'' $f_+$ and ``fraction of spins down'' $f_-$
are always important.
In the Ising-like model of protein folding problem, they correspond to the
``fraction of SUs folded'' and the ``faction of SUs unfolded,'' respectively.
They are related to $\avg{\sigma}$ by the simple relations
\begin{equation}
f_+=\frac{1+\avg{\sigma}}{2},\qquad f_-=\frac{1-\avg{\sigma}}{2}
\label{e:n011}
\end{equation}
These quantities are related to the observed ``fraction folded'' and ``fraction unfolded''
by Zwanzig.
In this work, this viewpoint is also adopted.
However, this is still debatable.
In a macroscopic experimental observation of the protein folding/unfolding process,
what can be said to be the native state of a protein?
When an enzyme is said to have 80 percent activity, for example, does it mean that
for each protein roughly around 80 percent of its  structural units are in their native states,
or does it mean that 80 percent of the protein molecules are almost perfectly folded,
while the states of the other 20 percent of molecules are not optimal enough?
To work on this abstract model, it is easier to accept the first viewpoint, 
but it is not clear which one is more reasonable.
This open problem cannot be treated in this work.

Finally, it should be noted that when $J=0$, 
$\avg{\sigma}_{\text{eq}}^*=\left(e^{2\bar{\epsilon}}-1\right)/
\left(e^{2\bar{\epsilon}}+1\right)=\tanh\bar{\epsilon}$,
where the superscripted star sign represents the ideal case throughout the following text.
For later convenience, the notation is further simplified so that 
$\sigma_{\infty}\equiv\avg{\sigma}_{\text{eq}}^*$.
In Eq.~\eqref{e:n010}, the terms involving $\bar{\epsilon}$ can be expressed in terms of 
$\sigma_{\infty}$.
Moreover, one can express $\sigma_{\infty}$ in terms of $\avg{\sigma}_{\text{eq}}$ and $\bar{J}$.
As a summary,
\begin{equation}
\avg{\sigma}_{\text{eq}}=\frac{\sigma_{\infty}e^{2\bar{J}}}{
\sqrt{1+\sigma_{\infty}\left(e^{4\bar{J}}-1\right)}};\qquad
\sigma_{\infty}=\frac{\avg{\sigma}_{\text{eq}}}{
\sqrt{\avg{\sigma}_{\text{eq}}^2+\left(1-\avg{\sigma}_{\text{eq}}^2\right)e^{4\bar{J}}}}
\label{e:n011a}
\end{equation}

\subsection{Kinetics}
The procedures for solving the kinetics of the present model are based on the
more general rate equation methods for treating general Ising models
presented by Glauber\cite{jmp4:294}.
This method is customized to solve the kinetics of exactly the present model
by Huang and Seitz\cite{cib:c3}.

To discuss the kinetics of the system, it is assumed that the elementary reactions are
the flippings of one spin at a time.
In reality, perhaps the simultaneous flipping of two or more spins is possible,
but in those cases it is more troublesome to construct the model phenomenologically.
In this work, especially, the principle of detailed balance is used to regulate the relative
magnitudes of the transition rates.
Given the equilibrium fraction of native and denatured SUs,
the ratio of forward and backward single-SU-flipping rate constants can be determined.
If it is desired to include double-SU-flipping processes,
the ratio of forward and backward double-SU-flipping rate constants have to be determined.
It will be seen later that in the present work the focus is on the quasi-chemical approximation
in which the ratio between number of neighboring SU-pair in difference configurations
are derived.
However, this is still not enough because double-SU-flipping processes are not limited
to happening to nearest-neighboring SU-pairs.
Presently, there does not seem to have a good principle for setting up
detail-balanced double-SU-flipping model.
Therefore, to avoid unwanted mistakes, those high-order processes are not considered.
Also noted is that, for brevity, in this paragraph \emph{flip} or \emph{flipping} were used
to describe the forward and backward transitions between the native and denatured
states of one SU.
These terms will be used throughout the text whenever there is no need to worry about ambiguity.

Starting with the initial spin configuration $\qket{\vec{\sigma}}$,
if the $j$-th SU flips, the value of $\sigma_j$ will change into $-\sigma_j$.
The resulted final structural configuration is different from $\qket{\vec{\sigma}}$ only by the
value of one SU.
A notation $\qket{\vec{\sigma}'_j}$ is introduced to denote this final state after the specific
process starting from $\qket{\vec{\sigma}}$.
The single-SU-flipping rates can be expressed as the matrix element of a
propagator $\hat{W}$, and the general master equation is%
\cite{jmp4:294,jsp29:375,pa137:439,pccp5:5300}
\begin{equation}
\frac{\rd}{\rd t}P\left(\vec{\sigma};t\right)=
-P\left(\vec{\sigma};t\right)\sum_j\qexp{\vec{\sigma}'_j}{\hat{W}}{\vec{\sigma}}
+\sum_jP\left(\vec{\sigma}'_j,t\right)\qexp{\vec{\sigma}}{\hat{W}}{\vec{\sigma}'_j}
\label{e:n012}
\end{equation}
where $P\left(\vec{\sigma};t\right)$ is the probability that the configuration of the system
is $\qket{\vec{\sigma}}$ at time $t$.
If $P\left(\vec{\sigma};t\right)$ is solved, one can obtain the kinetics of other
random variables of the system.
However, if the goal is to obtain the magnetization directly, there is a shortcut.
Since if $P\left(\vec{\sigma};t\right)$ is known, at time $t$ the expectation value of
$\sigma$, denoted simply as $\avg{\sigma}$, is
\begin{equation}
\avg{\sigma}=\sum_{\left\{\vec{\sigma}\right\}}\left[\frac{1}{M}\sum_j\sigma_j\right]
P\left(\vec{\sigma};t\right)
\label{e:n013}
\end{equation}
Differentiating $\avg{\sigma}$ with respect to time results in
\begin{equation}
\frac{\rd\avg{\sigma}}{\rd t}=
\sum_{\left\{\vec{\sigma}\right\}}\left[\frac{1}{M}\sum_j\sigma_j\right]
\frac{\rd P\left(\vec{\sigma};t\right)}{\rd t}
\label{e:n014}
\end{equation}
Inserting Eq.~\eqref{e:n012} into Eq.~\eqref{e:n014}, the right-hand-side can be easily simplified.
Since this is already worked out by Huang and Seitz\cite{cib:c3}, 
the derivation is left out.
Without any further approximation, the equation of motion (EOM) of the magnetization is
\begin{equation}
\frac{\rd\avg{\sigma}}{\rd t}=
\sum_{\left\{\vec{\sigma}\right\}}\frac{2}{M}\sum_j\sigma_j
\qexp{\vec{\sigma}'_j}{\hat{W}}{\vec{\sigma}}P\left(\vec{\sigma};t\right)
\label{e:n015}
\end{equation}
The principle of detailed balance, when applied to the Ising system, can be expressed in
the form introduced by Glauber\cite{jmp4:294}:
\begin{equation}
\frac{\qexp{\vec{\sigma}'_j}{\hat{W}}{\vec{\sigma}}}{\qexp{\vec{\sigma}}{\hat{W}}{\vec{\sigma}'_j}}
=\frac{\exp\left[-\bar{\epsilon}\sigma_j-\bar{J}\sigma_j
\left(\sigma_{j-1}+\sigma_{j+1}\right)\right]}{%
\exp\left[+\bar{\epsilon}\sigma_j+\bar{J}\sigma_j
\left(\sigma_{j-1}+\sigma_{j+1}\right)\right]}
\label{e:n016}
\end{equation}
By introducing a useful shorthand
$\eta=\tanh\left(2\bar{J}\right)$, It is found that
\begin{equation}
\frac{\qexp{\vec{\sigma}'_j}{\hat{W}}{\vec{\sigma}}}{\qexp{\vec{\sigma}}{\hat{W}}{\vec{\sigma}'_j}}
=\frac{1-\sigma_{\infty}\sigma_j}{1+\sigma_{\infty}\sigma_j}\times
\frac{1-\eta\sigma_j\left(\sigma_{j-1}+\sigma_{j+1}\right)/2}{%
1+\eta\sigma_j\left(\sigma_{j-1}+\sigma_{j+1}\right)/2}
\label{e:n017}
\end{equation}
Following the reasoning of Glauber, one can let
\begin{equation}
\qexp{\vec{\sigma}'_j}{\hat{W}}{\vec{\sigma}}=
\mathcal{W}\left(1-\sigma_{\infty}\sigma_j\right)
\left[1-\eta\sigma_j\left(\sigma_{j-1}+\sigma_{j+1}\right)/2\right]
\label{e:n018}
\end{equation}
where $\mathcal{W}$ is a characteristic rate (or frequency) of the system, or, in other words,
its inverse is a characteristic time scale of the system.
It has to be noted, however, that the Glauber's principle of detailed balance in the form of
Eq.~\eqref{e:n018} is worthy of further investigation.
Indeed, the meaning of this time scale and its connection with (or not with) the Markov
approximation were the topics of the study of Metiu~et~al.\cite{jcp63:5116}.
The problem not discussed in that work and other similar ones is that
if the proportional factor $\mathcal{W}$ also depends on the configuration of the system
but it does not depends on the sign of the $j$-th spin, condition in Eq.~\eqref{e:n016}
can still be satisfied, but the kinetics, if calculated by assuming Eq.~\eqref{e:n018} with
constant $\mathcal{W}$, may be unreliable.
This problem is beyond the scope of the present work.
In the following,
it is assumed that Eq.~\eqref{e:n018} with constant $\mathcal{W}$ is valid.
A deterministic EOM of the instantaneous expectation value $\avg{\sigma}$
can be obtained\cite{cib:c3}:
\begin{equation}
\frac{\rd\avg{\sigma}}{\rd\tau}=\sigma_{\infty}-\left(1-\eta\right)\avg{\sigma}
-\sigma_{\infty}\eta\avg{\frac{1}{M}\sum_j\sigma_j\sigma_{j+1}}
\label{e:n019}
\end{equation}
where $\tau=2\mathcal{W}t$ is the temporal variable $t$ scaled by the characteristic rate.

\subsection{Approximations}
The phenomenological equation \eqref{e:n019} cannot be solved analytically without
knowledge of the nearest-neighboring-SU correlation $\avg{\sum_j\sigma_j\sigma_{j+1}/M}$.
Actual knowledge about it does not exist, and approximation have to be made.

The first approximation, namely the mean-field approximation (MFA), 
is widely applied in literature\cite{pa137:439,jccs50:335,pccp5:5300}.
This is done by replacing the SU-SU correlation by $\avg{\sigma}^2$ so that
\begin{equation}
\frac{\rd\avg{\sigma}}{\rd\tau}=\sigma_{\infty}-\left(1-\eta\right)\avg{\sigma}
-\sigma_{\infty}\eta\avg{\sigma}^2
\label{e:a001}
\end{equation}
In this work the solution to this equation will not be discussed.
Nevertheless, the steady-state solution is examined.
By letting the time-derivative to be zero, it is found that at equilibrium the MFA model
predicts that
\begin{equation}
\avg{\sigma}_{\text{eq(MFA)}}=\frac{\eta-1+\sqrt{\eta^2+\left(4\sigma_{\infty}^2-2\right)\eta+1}}{2\sigma_{\infty}\eta}
\label{e:a002}
\end{equation}
In comparison, expressed in terms of $\sigma_{\infty}$ and $\eta$, the exact solution 
Eq.~\eqref{e:n010} becomes
\begin{equation}
\avg{\sigma}_{\text{eq}}=\frac{\sigma_{\infty}\sqrt{1+\eta}}{\sqrt{1-\eta+2\eta\sigma_{\infty}^2}}
\label{e:a003}
\end{equation}
Thus, MFA predicts wrong equilibrium solution, 
although typically the numerical difference is not serious.
This is why the kinetics of MFA model is not of interests here.

The second and the major approximation to be discussed in the following is the
quasi-chemical approximation (QCA)\cite{cib:c3,nature148:304,tfs44:1007}.
The details of the application of QCA to the present model
are fully explained by Huang and Seitz\cite{cib:c3}.
In the following, only the outline of the method is presented.

For an arbitrary configuration of the system (of $M$ SUs),
let the number of native SUs be $M_+$, and the number of denatured SUs be $M_-$.
Meanwhile, the number of neighboring pairs that are both native is $M_{++}$ and the number
of neighboring pairs that are both dnatured is $M_{--}$.
In parallel, the number of neighboring pairs with opposite states is $M_{+-}$.
For very large $M$, ignore the boundary effect by assuming cyclic boundary conditions and
it can be shown that $2M_+=2M_{++}+M_{+-}$ and $2M_-=2M_{--}+M_{+-}$ and 
$M_-=M-M_+$.
In the summation $\sum_j\sigma_j\sigma_{j+1}$ in the SU-SU correlation,
the $M_{++}$ pairs and $M_{--}$ pairs of spins each contributes $+1$,
and the $M_{+-}$ pairs of spins each contributes $-1$ to the result:
\begin{equation}
\frac{1}{M}\sum_j\sigma_j\sigma_{j+1}=\frac{M_{++}+M_{--}-M_{+-}}{M}
=\frac{4M_{++}-4M_++M}{M}
\label{e:a004}
\end{equation}
Therefore if $\avg{M_{++}}$ is found, the expected SU-SU correlation can also be found.
The necessity of an approximation lies in the lack of a self-consistent knowledge of
the probability $P\left(\vec{\sigma};\tau\right)$ for evaluating the average.
In QCA it is proposed that, knowing that the \emph{local} energies of
a ++, +$-$/$-$+ and $-$$-$ pair of spins are, 
respectively, $-2\epsilon-J$, $J$ and $2\epsilon-J$,
the numbers of these kinds of pairs are distributed, at any moment, according to the
their respective Boltzmann factors.
In other words, $\avg{M_{+-}}/\avg{M_{++}}=e^{-\bar{J}}/e^{2\bar{\epsilon}+\bar{J}}$,
$\avg{M_{+-}}/\avg{M_{--}}=e^{-\bar{J}}/e^{-2\bar{\epsilon}+\bar{J}}$, and therefore
\begin{equation}
\frac{\avg{M_{+-}}^2}{\avg{M_{++}}\avg{M_{--}}}=
\frac{e^{-2\bar{J}}}{e^{2\bar{\epsilon}+\bar{J}}e^{-2\bar{\epsilon}+\bar{J}}}=e^{-4\bar{J}}
\label{e:a005}
\end{equation}
A few more algebraic steps should be carried out to convert this condition into an expression of
$\avg{M_{++}}$ as a function of $\avg{M_+}$ (and the constant $\bar{J}$) only.
Importantly, the fraction of native SUs, $f_+=\left(1+\avg{\sigma}\right)/2$
is directly linked to $\avg{M_+}$ by $f_+=\avg{M_+}/M$ so
\begin{equation}
\frac{\avg{M_{++}}}{M}=\frac{\avg{\sigma}+1}{2}
\left[1+\frac{\avg{\sigma}-1}{1+
\sqrt{\avg{\sigma}^2+\left(1-\avg{\sigma}^2\right)e^{4\bar{J}}}}\right]
\label{e:a006}
\end{equation}
and the EOM of $\avg{\sigma}$ becomes
\begin{equation}
\frac{\rd\avg{\sigma}}{\rd\tau}=-\left(1-\eta\right)
\left[\avg{\sigma}-
\sigma_{\infty}\sqrt{\avg{\sigma}^2+\left(1-\avg{\sigma}^2\right)e^{4\bar{J}}}\right]
\label{e:a007}
\end{equation}
The steady-state solution of this equation is especially interesting.
If $\rd\avg{\sigma}/\rd\tau=0$, and call the value of $\avg{\sigma}$
in this steady state $\avg{\sigma}_{\text{eq}}$,
it is found that
\begin{equation}
\sigma_{\infty}=\frac{\avg{\sigma}_{\text{eq}}}{
\sqrt{\avg{\sigma}_{\text{eq}}^2+\left(1-\avg{\sigma}_{\text{eq}}^2\right)e^{4\bar{J}}}}
\label{e:a008}
\end{equation}
which is exactly the equilibrium solution of the model system, Eq.~\eqref{e:n011a}.

It is instructive to emphasize the difference between MFA and QCA again.
In both approximations, certain quantities are assumed to be at their steady-state values.
In MFA, the steady-state fractions of native and denatured SUs that are used to
estimate the SU-SU correlation.
In QCA, the steady-state fractions of ++, +$-$/$-$+ and $-$$-$ spin-pairs are used instead.
Unless the SUs are all independent, MFA can never predict the correct equilibrium analytically.
When there are couplings between spins, QCA catches up the lowest-order effect.
If there were long-range couplings, QCA cannot be exact analytically, either.
In our constant-NNI model, however, the range of interaction matches that assumed by QCA,
and the correct equilibrium properties of the system can be obtained.
When working on the kinetics, it is therefore also expected that adopting QCA is exactly
equivalent to applying steady-state approximation, although this is not proved.

Analytic expression of the solution of Eq.~\eqref{e:a007} is available\cite{cib:c3},
but the explicit form is not of interests in the present work.
Instead, in the next Section the alternative form of the EOM, Eq.~\eqref{e:a007}, 
will be discussed.

\section{Results and Discussions}
In this Section, the new contributions of the author in this work are summarized.

Consider the reaction of a protein from its native state (N) to the denatured state (D):
\begin{equation}
{\rm N} \mathrel{\mathop{\rightleftarrows}^{K}} {\rm D}
\label{e:d001}
\end{equation}
In the present model, if the coupling $J$ is zero, each SU flips independently, and the
equilibrium is determined by $\epsilon$.
This case shall be referred to as the \emph{ideal} case in the following.
The equilibrium constant also has an ideal value: $K^*=\exp\left(-2\bar{\epsilon}\right)$.
In terms of fractions of spins up or down, 
or in terms of the magnetization, this equilibrium is expressed as
\begin{gather}
\frac{f_{-\text{,eq}}^*}{f_{+\text{,eq}}^*}=
\frac{1-\avg{\sigma}_{\text{eq}}^*}{1+\avg{\sigma}_{\text{eq}}^*}=K^*=e^{-2\bar{\epsilon}}
\label{e:d002}\\
\avg{\sigma}_{\text{eq}}^*=\frac{e^{2\bar{\epsilon}}-1}{e^{2\bar{\epsilon}}+1}
\label{e:d003}
\end{gather}
Notice that $\epsilon$ is not determined solely by the intrinsic properties of the system.
It depends on external conditions.

When $J\ne0$, the equilibrium solution is obviously different from the solution in the
ideal case.
However, the free-energy change upon unfolding is the same.
When the SUs are all in the folded state $\sigma=1$, the total free energy of the system
with $M$ SUs is $M\left(-\epsilon-J\right)$.
In other words, the free-energy per SU is $-\epsilon-J$.
Similarly, the free-energy per SU when the fraction unfolded in 1 is $\epsilon-J$.
In other words, the denaturation reaction free energy change is 
$\Delta_{\rm d}G=\epsilon-J-\epsilon+J=2\epsilon$. 
The equilibium constant, then, is
$K=\exp\left[-\Delta_{\rm d}G/\kBT\right]
=\exp\left(-2\bar{\epsilon}\right)=K^*$.
However, the ratio between the equilibrium values of
the fractions $f_{-\text{,eq}}$ and $f_{+\text{,eq}}$ is
\begin{equation}
\frac{f_{-\text{,eq}}}{f_{+\text{,eq}}}=\frac{
\sqrt{\left(e^{2\bar{\epsilon}}-1\right)^2+2e^{2\bar{\epsilon}}e^{-4\bar{J}}}-
\left(e^{2\bar{\epsilon}}-1\right)}{
\sqrt{\left(e^{2\bar{\epsilon}}-1\right)^2+2e^{2\bar{\epsilon}}e^{-4\bar{J}}}+
\left(e^{2\bar{\epsilon}}-1\right)}
\label{e:d004}
\end{equation}
which is obviously different from $K$.
This is not surprising because with the non-ideality $J$,
it is expected that the activity coefficients $\gamma_+$ and $\gamma_-$
together with the activities $a_+$ and $a_-$ should be defined so that
\begin{equation}
a_+=\gamma_+f_{+\text{,eq}};\qquad
a_-=\gamma_-f_{-\text{,eq}}
\label{e:d005}\end{equation}
and
\begin{equation}
\frac{a_-}{a_+}=\frac{\gamma_-f_{-\text{,eq}}}{\gamma_+f_{+\text{f,eq}}}
=K=e^{-2\bar{\epsilon}}
\label{e:d006}
\end{equation}
With some rearrangement, it is found that
$\gamma_+$ and $\gamma_-$ has to satisfy the relations
\begin{equation}
\frac{\gamma_-}{\gamma_+}=
\frac{f_{+\text{,eq}}}{f_{-\text{,eq}}}\cdot e^{-2\bar{\epsilon}}=
\frac{\left(\sqrt{\avg{\sigma}_{\text{eq}}^2+e^{4\bar{J}}\left(1-\avg{\sigma}_{\text{eq}}^2\right)}
-\avg{\sigma}_{\text{eq}}\right)/\left(1-\avg{\sigma}_{\text{eq}}\right)}{
\left(\sqrt{\avg{\sigma}_{\text{eq}}^2+e^{4\bar{J}}\left(1-\avg{\sigma}_{\text{eq}}^2\right)}
+\avg{\sigma}_{\text{eq}}\right)/\left(1+\avg{\sigma}_{\text{eq}}\right)}
\label{e:d007}
\end{equation}
and $\lim_{J\rightarrow0}\gamma_{\pm}=1$, and 
$\lim_{J\rightarrow0}a_{\pm}=f_{\pm,\text{eq}}^*$.
The seemingly most reasonable choice is that
\begin{equation}
\gamma_{\pm}=\frac{1\pm\avg{\sigma}_{\text{eq}}/
\sqrt{\avg{\sigma}_{\text{eq}}^2+e^{4\bar{J}}\left(1-\avg{\sigma}_{\text{eq}}^2\right)}
}{1\pm\avg{\sigma}_{\text{eq}}}
\label{e:d008}
\end{equation}
Consequently
\begin{equation}
a_{\pm}=\frac{1\pm\avg{\sigma}_{\text{eq}}/
\sqrt{\avg{\sigma}_{\text{eq}}^2+e^{4\bar{J}}\left(1-\avg{\sigma}_{\text{eq}}^2\right)}
}{2}
\label{e:d009}
\end{equation}
However, even with the constraints listed above, the choice of the form of the activities and
coefficients are not unique.
For example, an extra factor of $2/\left(1+\exp\left(4\bar{J}\right)\right)$ can
be included and all of the constraints can still be satisfied.
Anyway, with the above choice it is also quite straightforward to define
an activity coefficient and activity for $\avg{\sigma}$:
\begin{equation}
\gamma_{\sigma}=\frac{1}
{\sqrt{\avg{\sigma}_{\text{eq}}^2+e^{4\bar{J}}\left(1-\avg{\sigma}_{\text{eq}}^2\right)}};\qquad
a_{\sigma}=\frac{\avg{\sigma}_{\text{eq}}}
{\sqrt{\avg{\sigma}_{\text{eq}}^2+e^{4\bar{J}}\left(1-\avg{\sigma}_{\text{eq}}^2\right)}}
\label{e:d010}
\end{equation}
Based on wild imagination, the author decided to define several quantities.
In parallel with the activities and activity coefficients, the kinetic versions of these
quantities, or \emph{instantaneous activities} $\alpha_{\pm}$ and 
\emph{instantaneous activity coefficients} $\Gamma_{\pm}$ are defined as
\begin{equation}
\begin{aligned}
\Gamma_{\pm}&=\frac{1\pm\avg{\sigma}/
\sqrt{\avg{\sigma}^2+e^{4\bar{J}}\left(1-\avg{\sigma}^2\right)}
}{1\pm\avg{\sigma}} 
\\ 
\alpha_{\pm}&=\frac{1\pm\avg{\sigma}/
\sqrt{\avg{\sigma}^2+e^{4\bar{J}}\left(1-\avg{\sigma}^2\right)}}{2}
\end{aligned}
\label{e:d011}
\end{equation}
It is also straightforward that
\begin{equation}
\Gamma_{\sigma}=\frac{1}
{\sqrt{\avg{\sigma}^2+e^{4\bar{J}}\left(1-\avg{\sigma}^2\right)}};\qquad
\alpha_{\sigma}=\frac{\avg{\sigma}}
{\sqrt{\avg{\sigma}^2+e^{4\bar{J}}\left(1-\avg{\sigma}^2\right)}}
\label{e:d013}
\end{equation}
Finally, notice the relation:
\begin{equation}
\sigma_{\infty}=a_{\sigma}
\label{e:d014}
\end{equation}
With the above definitions and relations, the EOM Eq.\eqref{e:a007} becomes
\begin{align}
\frac{\rd\avg{\sigma}}{\rd\tau}&=-\left(1-\eta\right)\left(
\avg{\sigma}-
a_{\sigma}\sqrt{\avg{\sigma}^2+e^{4\bar{J}}\left(1-\avg{\sigma}^2\right)}\right)\nonumber\\
&=-\frac{2}{1+e^{4\bar{J}}}\left(\frac{\alpha_{\sigma}}{\Gamma_{\sigma}}-
\frac{a_{\sigma}}{\Gamma_{\sigma}}\right)=
-\frac{2}{1+e^{4\bar{J}}}\,\frac{\alpha_{\sigma}-a_{\sigma}}{\Gamma_{\sigma}}
\label{e:d015}
\end{align}
It can also be shown that
\begin{equation}
\frac{\rd\alpha_{\sigma}}{\rd\tau}=\frac{\rd\left(\alpha_{\sigma}-a_{\sigma}\right)}{\rd\tau}=
-\frac{2\Gamma_{\sigma}^2}{1+e^{-4\bar{J}}}
\left(\alpha_{\sigma}-a_{\sigma}\right)
\label{e:d016}
\end{equation}
These equations, never derived in such ways in the literature, have interesting but actually
not surprising structures.
Take Eq.\eqref{e:d016} as example, it indicates that the deviation of the `instantaneous'
activity from its equilibrium value relaxes in an almost first-order manner.
However, the factor $2\Gamma_{\sigma}^2/\left[1+\exp\left(-4\bar{J}\right)\right]$,
which takes the place of a rate constant of a first-order reaction,
is time-dependent through $\Gamma_{\sigma}$.
This \emph{rate factor} goes to unity when the cooperativity $J$ goes to zero.
In that case, $\alpha_{\sigma}$ approaches $\avg{\sigma}$ and $a_{\sigma}$
approaches $\avg{\sigma}_{\text{eq}}$.
The reaction rate is determined solely by the characteristic rate $\mathcal{W}$,
as expected.

\begin{figure}[ht]
\begin{center}
\includegraphics[width=12cm]{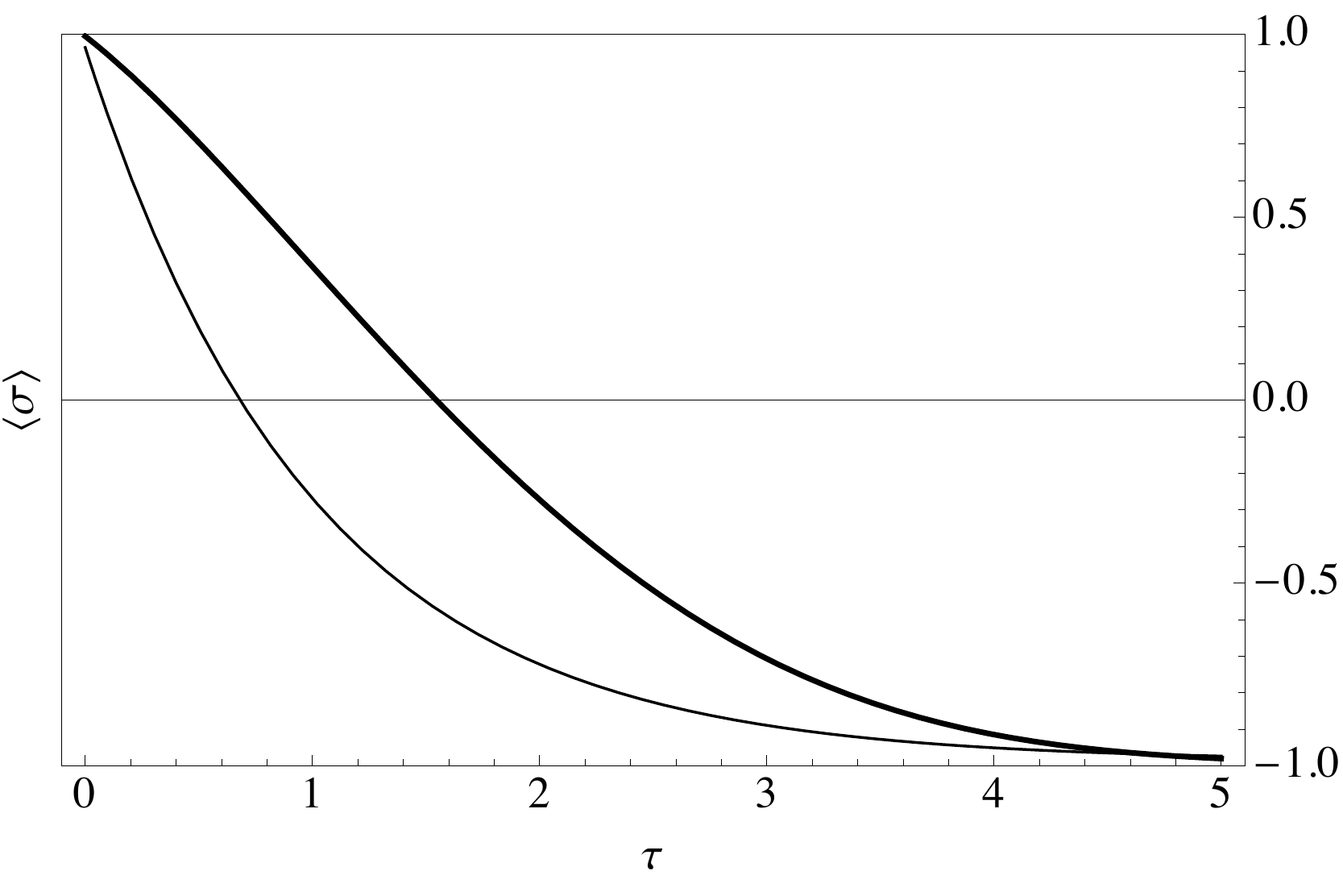}
\end{center}
\caption{The denaturing kinetics for a model system. 
Initially the external conditions make the local free energy difference $\epsilon=2\kBT$.
At the moment when the scaled time $\tau=0$, 
the external conditions are suddenly changed so that
the local free energy difference abruptly changes into $\epsilon=-2.5\kBT$.
The thinner solid line depicts the case when $J=0$, and the thick solid line
corresponds to the case in which $J=0.5\kBT$.
}\label{f:sigmataujp}
\end{figure}
\begin{figure}[ht]
\begin{center}
\includegraphics[width=12cm]{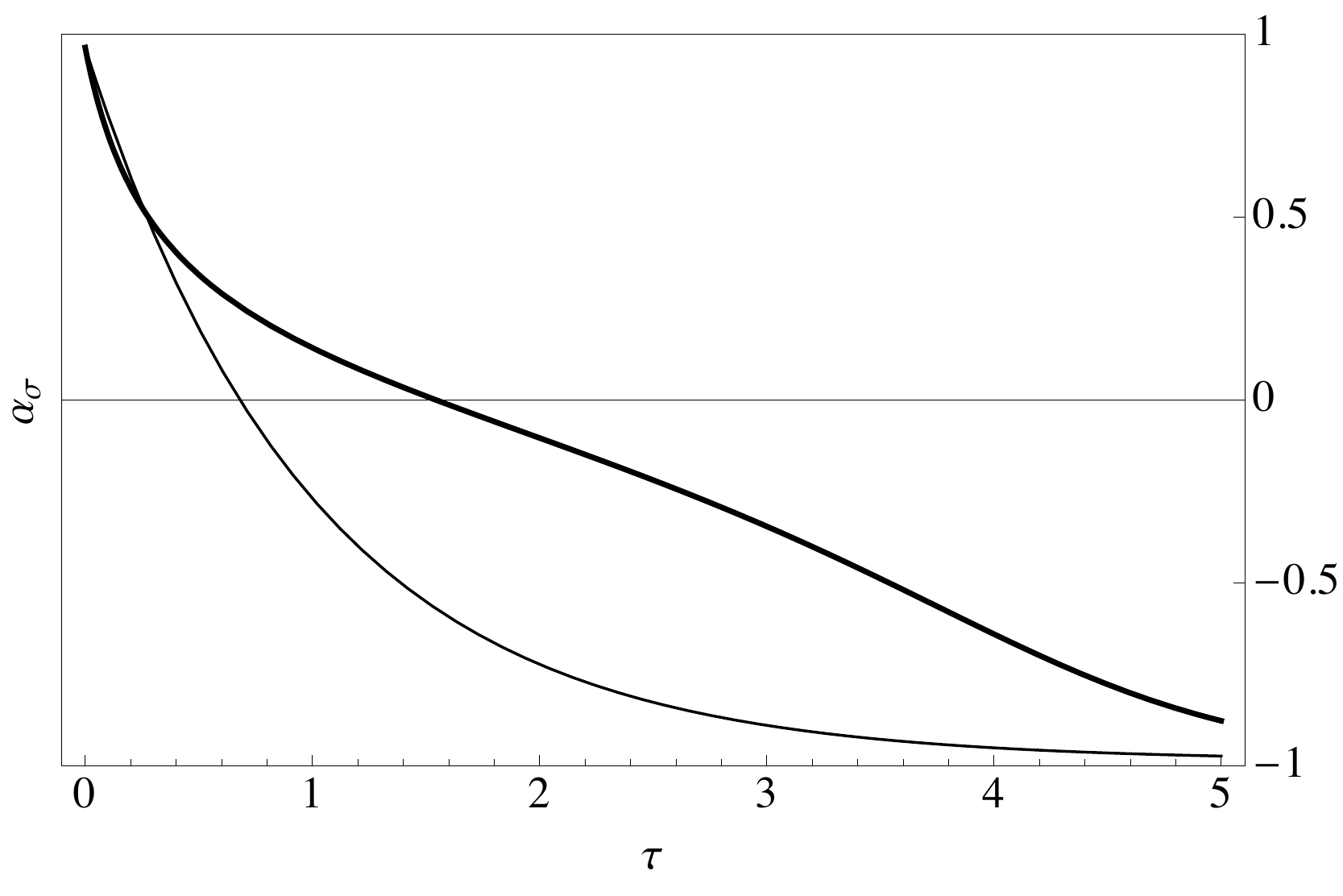}
\end{center}
\caption{The denaturing kinetics of the same model system as that shown in 
Fig.~(\protect{\ref{f:sigmataujp}}).
The instantaneous activities are plotted against time.
The thinner line depicts the non-cooperative case, while the thick line corresponds to
the case in which $J=0.5\kBT$.}
\label{f:alphataujp}
\end{figure}
\begin{figure}[ht]
\begin{center}
\includegraphics[width=12cm]{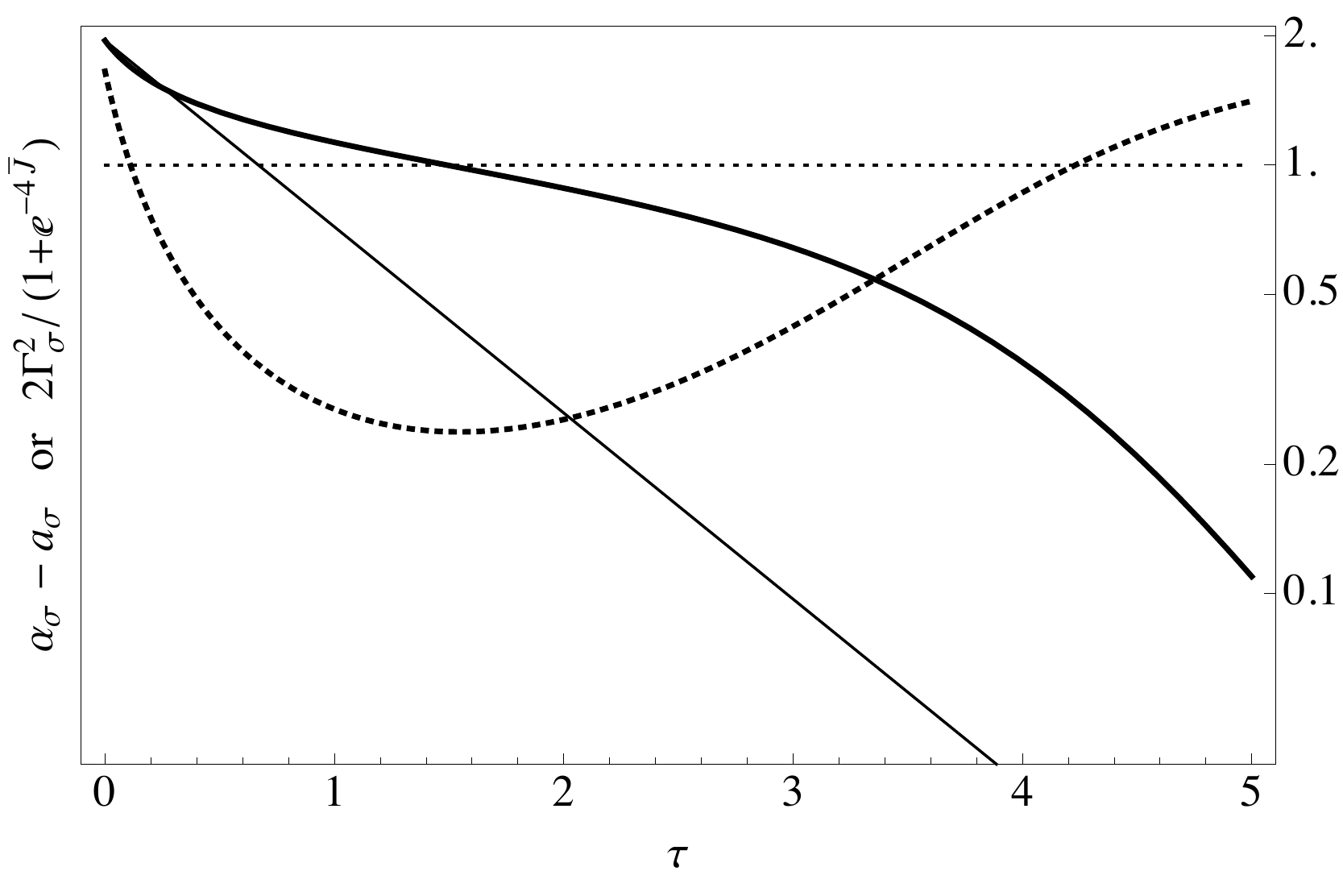}
\end{center}
\caption{The denaturing kinetics of the same model system as that shown in 
Fig.~(\protect{\ref{f:sigmataujp}}).
The thinner lines depict the non-cooperative case, while the thick lines correspond to
the case in which $J=0.5\kBT$.
The solid lines are the instantaneous activities subtracted by the equilibrium activities, 
and the dotted lines are the rate factors $2\Gamma_{\sigma}^2/\left(1+e^{4\bar{J}}\right)$.
}\label{f:logalphataujp}
\end{figure}

In order to understand better the cooperative effects with the help of the above expressions,
in Fig.~(\ref{f:sigmataujp}) through (\ref{f:logalphataujp}), 
the unfolding kinetics of a non-cooperative $\left(J=0\right)$ system
and a positively cooperative system
$\left(J=0.5\kBT>0\right)$ are shown in three different ways.
The system is first `folded' by setting its initial local free energy difference to $\epsilon=2\kBT$.
At the moment when the scaled time $\tau=0$, 
the external conditions are suddenly changed so that
the local free energy difference abruptly changes into $\epsilon=-2.5\kBT$.
In Fig.~(\ref{f:sigmataujp}), the `magnetization' $\avg{\sigma}$ is plotted against $\tau$.
The thinner line depicts the non-cooperative kinetics, while the thick line
is the cooperative kinetics.
The initial values of $\avg{\sigma}$ are a bit different due to different values of $J$.
Then, the cooperative system relaxes much more slowly.
If the fraction folded is plotted instead, the plots are simply shifted and rescaled,
and qualitatively they look all the same as this figure.
It is interesting to see that positive cooperativity slows down the kinetics.
The reason will be discussed later.
To see the trends more clearly, in Fig.~(\ref{f:alphataujp}) the instantaneous activities
of both the non-cooperative (thinner line) and the positively cooperative systems are shown.
The initial instantaneous activities values of both systems are the same.
The thick line which represents the kinetics of the positively cooperative system
obviously shows a slower kinetics, except for a short period at the beginning.
After $\tau=3$, however, the cooperative system seems to speed up its denaturing process.
It should be mentioned at this stage that, if a renaturation process is simulated instead,
the trend is also the same (not shown).
In other words, the unfolding process slows down in positively cooperative system
not because the native state is favored by positive cooperativity.
To find out the reason behind this feature, in Fig.~(\ref{f:logalphataujp}),
the difference between the instantaneous activities and the equilibrium activities 
of both non-cooperative and cooperative cases are plotted in logarithmic scale.
They are the thin and thick solid lines, respectively.
Also plotted in the same figure are the rate factor 
$2\Gamma_{\sigma}^2/\left(1+e^{-4\bar{J}}\right)$, in dotted lines.
In the non-cooperative case, the difference $\alpha_{\sigma}-a_{\sigma}$
relaxes in single-exponential manner, and the rate factor is unity.
This clearly reflects the two-state character of the system.
A collection of independent two-level systems behaves just like one two-state system.
In the positively cooperative case, in contrast, in a short period at the beginning,
the rate factor is greater than unity.
However, very quickly this rate factor reduces below unity.
This is because when $\avg{\sigma}$ decreases from the positive initial value to a
value closer to zero, the numbers of native and unfolded SUs are close to each other,
and number of neighboring SUs with opposite states is relatively high.
In that case, whether or not the flipping of one of the SUs is
energetically favored is less certain.
In other words, the system is trapped in a \emph{frustrated} state.
Only until the number of native SUs reduces to a rather low level,
the unfolding process becomes not so frustrated.
The rate factor increases to above unity again, and the denaturing kinetics becomes
faster than in the non-cooperative case.

Since
\begin{equation}
\frac{\rd\avg{\sigma}_{\text{eq}}}{\rd \bar{J}}=
\frac{2e^{2\bar{J}}\sinh\bar{\epsilon}}{\left(1+e^{4\bar{J}}\sinh^2\bar{\epsilon}\right)^{3/2}}
\label{e:d017}
\end{equation}
is positive when $\bar{\epsilon}>0$ and negative when $\bar{\epsilon}<0$,
positive cooperativity will result in higher fraction folded when the external conditions
favor folding, and will result in higher fraction unfolded when the external conditions
favor unfolding.
It would have been expected that positive cooperativity will also 
enhance the refolding/unfolding kinetics.
However, in reality, the positive cooperativity introduces a frustrating stage 
half-way in the refolding/unfolding processes.
Nevertheless, this frustrated phase will eventually pass and 
the system will approach its equilibrium in a faster rate in the later stage.
It is worthy of further studies that, if a model can be established in which
the coupling strength $J$ is different between native-native pair and 
denatured-denatured pair of SUs, will this frustration phenomenon change?
This will be the topic of later studies.

\begin{figure}[ht]
\begin{center}
\includegraphics[width=12cm]{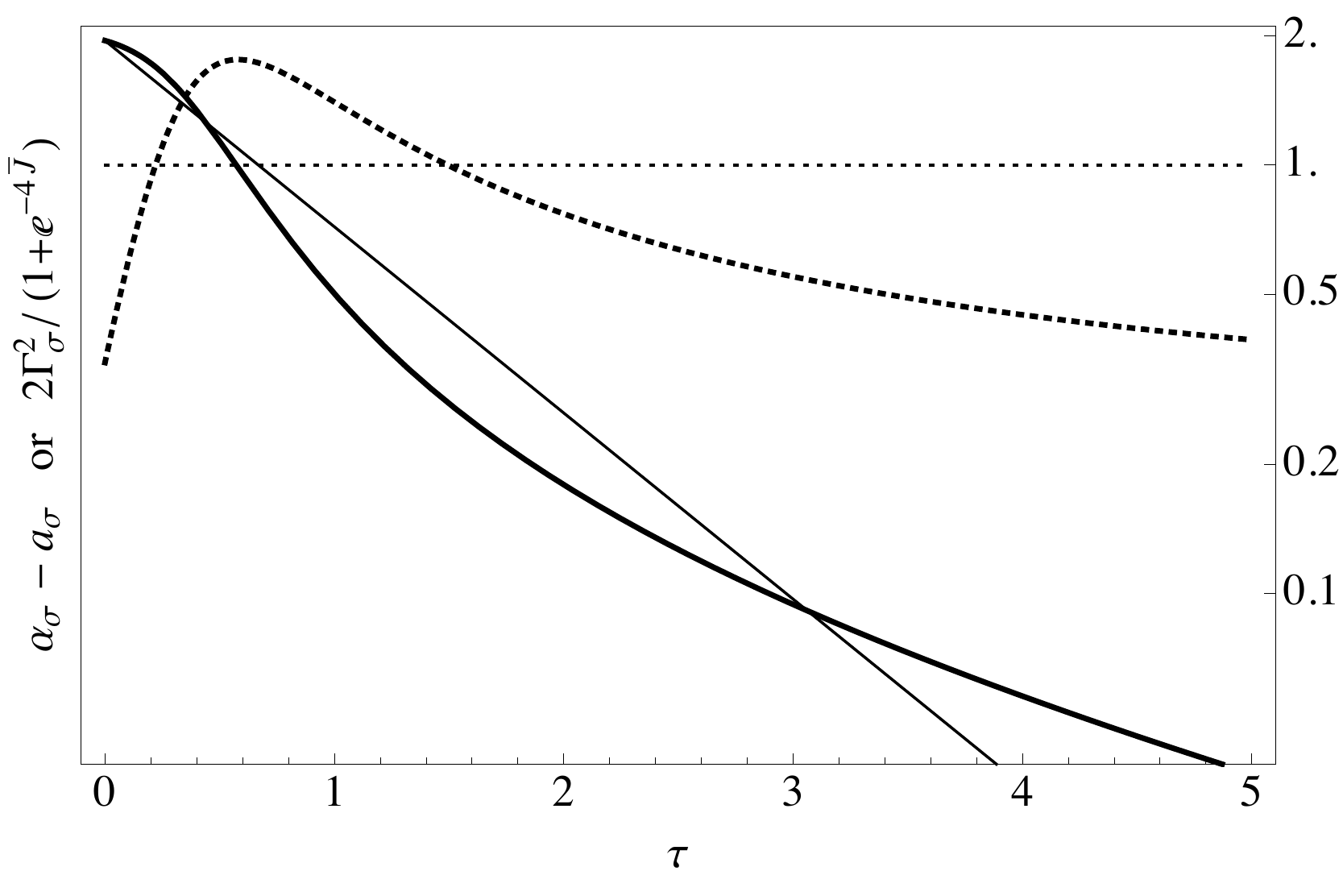}
\end{center}
\caption{The denaturing kinetics of a similar model system as that shown in 
Fig.~(\protect{\ref{f:sigmataujp}}).
The thinner lines depict the non-cooperative case, while the thick lines correspond to
the case in which $J=-0.5\kBT$.
The solid lines are the instantaneous activities subtracted by the equilibrium activities, 
and the dotted lines are the rate factors $2\Gamma_{\sigma}^2/\left(1+e^{4\bar{J}}\right)$.
}\label{f:logalphataujm}
\end{figure}
The behavior of negatively cooperative system can also be understood now.
In Fig.~(\ref{f:logalphataujm}), the differences between the instantaneous activities and
the equilibrium activities, $\alpha_{\sigma}-a_{\sigma}$, for the non-cooperative system and
a negatively cooperative system, are plotted.
The rate factors in both cases are also plotted.
At the very beginning and somewhat later, the rate factor of the negatively-cooperative system
is less than unity.
Only for a rather short period of time at the early stage,
the rate factor of the negatively cooperative system is greater than unity.
It seems that most of the time the system is quite frustrated.
In the short period when the unfolding process seems fast,
there are too many SUs in native state and they are forced to transit.
But this intermediate stage does not last long.

Indeed, the behavior of this model system shows interesting features,
but is it a coincidence of a special model system, or does it provide useful insight to
what would be expected with other model systems?
It was shown in Eq.~\eqref{e:a008} that the steady-state solution of the EOM of QCKI model
is the same as the exact equilibrium solution of the constant-NNI Ising model,
and the reason is also discussed there.
The implication of this result is that,
by first applying the Glauber's principle of detail balance,
and next assuming that the SU-SU correlation can be replaced by its equilibrium value
under the given value of $\avg{\sigma}$,
it is equivalent to saying that at any moment during the process
$\rd\avg{\sigma}/\rd\tau$ is about zero.
In other words, for the constant-NNI Ising model, detailed balance and
QCA guarantees the steady-state approximation (SSA), and vise versa.
Moreover, it is implied that the function forms of the equilibrium activity $a_{\sigma}$
and the activity coefficient $\gamma_{\sigma}$ are very special function forms of
the instantaneous $\avg{\sigma}$.
Although in this work no attempt was made to prove the generality of this feature,
it seems reasonable that, for any kind of Ising-like models,
if efforts can be made to find out the function form of the activity and activity coefficient,
the steady-state kinetics can be expressed in terms of the instantaneous version of
these important functions.

Steady-state approximation is one of the theoretical cornerstones of the field of enzyme kinetics.
Although, for individual systems, it is quite difficult to analyze how good the SSA can apply,
most of the techniques for interpreting experimental data were based on SSA.
In the following, the chevron plot of the refolding kinetics of the 
1-D constant-NNI Ising model will be discussed to reveal further insights into the
cooperative effects.

The chevron plot is the plot of the logarithmic of the observed rate constant of a
refolding/unfolding reaction, as a function of a variety of variables\cite{jmb15:489}.
Among others, denaturant concentration is one of the most often seen variables.
In the present model, the denaturant concentration is not introduced.
However, the conceptual basis behind the interpretation of the chevron plot is the
empirical rule that the logarithm of the
forward and backward (renaturation and denaturation) reaction rate constants
depends linearly on the denaturant concentration\cite{jmb15:489}.
From the thermodynamic viewpoint, this trend means that
the activation energies for both the forward and backward reactions
depend on the denaturant concentration in a simple power-law form.
Since only (nearly) two-state proteins seem to have one better-defined transition state,
this concept is easier to apply to two-state proteins.
For a reaction with more intermediate states, if there is still one well-defined reaction path,
the theory behind chevron plot can still be applied, with some modifications to
match the complexity of the problem.
However, if there is not one dominant reaction path, and the funnel concept has to be applied, 
the application of chevron plot theory and transition state theory 
become not so manageable.

In the present model, there is an advantage that only a set of weakly-coupled 
local two-state systems is discussed.
Although the inter-SU couplings made the energetics more complicated than a two-state system,
the present model is still rather close to, and in the limiting case goes to, the two-state system.
In other words, as the lowest-order approximation, 
it can be assumed that there is only one transition state between the native and unfolded states.
Stated in a more abstract way, it can be assumed that one representative value of the 
free energy of the transition state can be used to calculate the activation energies
(and therefore the rate constants) of both the forward and the backward reactions.
In that case, apart from a multiplicative constant, the observed rate constant
can be determined from the reaction free-energy change and the activity coefficients.
The activity coefficients have to come into the expressions of the rate constants because
the determinations of the rate constants are based on the functional relation of the
reaction rate with the concentrations of the reactants.
In contrast, the reaction free-energy change is a function of the standard free-energy change
and the activities of the reactants.
When non-ideality is the focus of investigation, the role of the activity coefficients
cannot be neglected.
In the following $k_+$ and $k_-$ are defined as the rate constants of refolding
and unfolding reactions, separately,
but in the present model, they are not really constants.
From Eq.~\eqref{e:d015}, considering that $\alpha_{\sigma}=\alpha_+-\alpha_-$,
$a_{\sigma}=a_+-a_-$, and $\rd\sigma/\rd\tau=\pm2\rd f_{\pm}/\rd\tau$,
it is found that
\begin{equation}
\frac{\rd f_{\pm}}{\rd\tau}=\mp\frac{1}{1+e^{4\bar{J}}}
\frac{\left(\Gamma_+f_+-a_+\right)-\left(\Gamma_-f_--a_-\right)}{\Gamma_{\sigma}}
\label{e:d018}
\end{equation}
The expression suggests that, besides a constant multiplicative factor $2\mathcal{W}$, 
in the place of $k_+$ and $k_-$  there are
\begin{equation}
\begin{aligned}
k_+&=\frac{\Gamma_+}{\left(1+e^{4\bar{J}}\right)\Gamma_{\sigma}}
=\frac{\sqrt{\avg{\sigma}^2+e^{4\bar{J}}\left(1-\avg{\sigma}^2\right)}+\avg{\sigma}}{
\left(1+e^{4\bar{J}}\right)\left(1+\avg{\sigma}\right)}
\\ 
k_-&=\frac{\Gamma_-}{\left(1+e^{4\bar{J}}\right)\Gamma_{\sigma}}
=\frac{\sqrt{\avg{\sigma}^2+e^{4\bar{J}}\left(1-\avg{\sigma}^2\right)}-\avg{\sigma}}{
\left(1+e^{4\bar{J}}\right)\left(1-\avg{\sigma}\right)}
\end{aligned}
\label{e:d019}
\end{equation}
Consequently,
\begin{equation}
k_{\text{obs}}=k_++k_-
=2\frac{\sqrt{\avg{\sigma}^2+e^{4\bar{J}}\left(1-\avg{\sigma}^2\right)}-\avg{\sigma}^2}{
\left(1+e^{4\bar{J}}\right)\left(1-\avg{\sigma}^2\right)}
\label{e:d020}
\end{equation}

In this model, the external conditions that can affect the value of $\epsilon$ is not discussed.
Therefore, the better way to make the chevron plot is to plot $\log k_{\text{obs}}$ against
$\bar{\epsilon}$.
However, in Eq.~\eqref{e:d020}, $k_{\text{obs}}$ is expressed in terms of $\avg{\sigma}$
and $\bar{J}$.
This is for emphasizing that the reaction rate constant is actually not a constant, 
and it depends on the instantaneous configuration of the system, represented by $\avg{\sigma}$.
But to make the chevron plot, it is necessary to determine a value of $\epsilon$.
The strategy used here is to imagine that experimentally the initial-rate method is used 
to determine $k_{\text{obs}}$.
That is, imagine that when conducting the unfolding experiment
the system is first prepared in a condition in which the value of the local free energy difference
is $\epsilon$, and the average `magnetization' of this system, $\avg{\sigma}$,
takes the equilibrium value under $\epsilon$ through Eq.~\eqref{e:n010}.
At time $\tau=0$, the relaxation process is initiated by changing the external conditions.
Then the initial rate constant is measured.
Under these assumptions,
the initial rate constant should be close to that predicted by the present theory.
It is worthy of noting that the expression of $k_{\text{obs}}$, Eq.~\eqref{e:d020},
does not depend on the final value of $\epsilon$ in the relaxation process.
This suggests that the expression of $k_{\text{obs}}$ is instantaneous.
Unfortunately, most of the experiments that determined the chevron plots of certain proteins
were not performed with the initial-rate method.
The fraction unfolded changes over as much as two orders of magnitude in many cases.
However, in the present work there is not a theory to simulate those kind of experiments.
The discussions below will be limited to the initial rates predicted by the present model.

\begin{figure}[ht]
\begin{center}
\includegraphics[width=12cm]{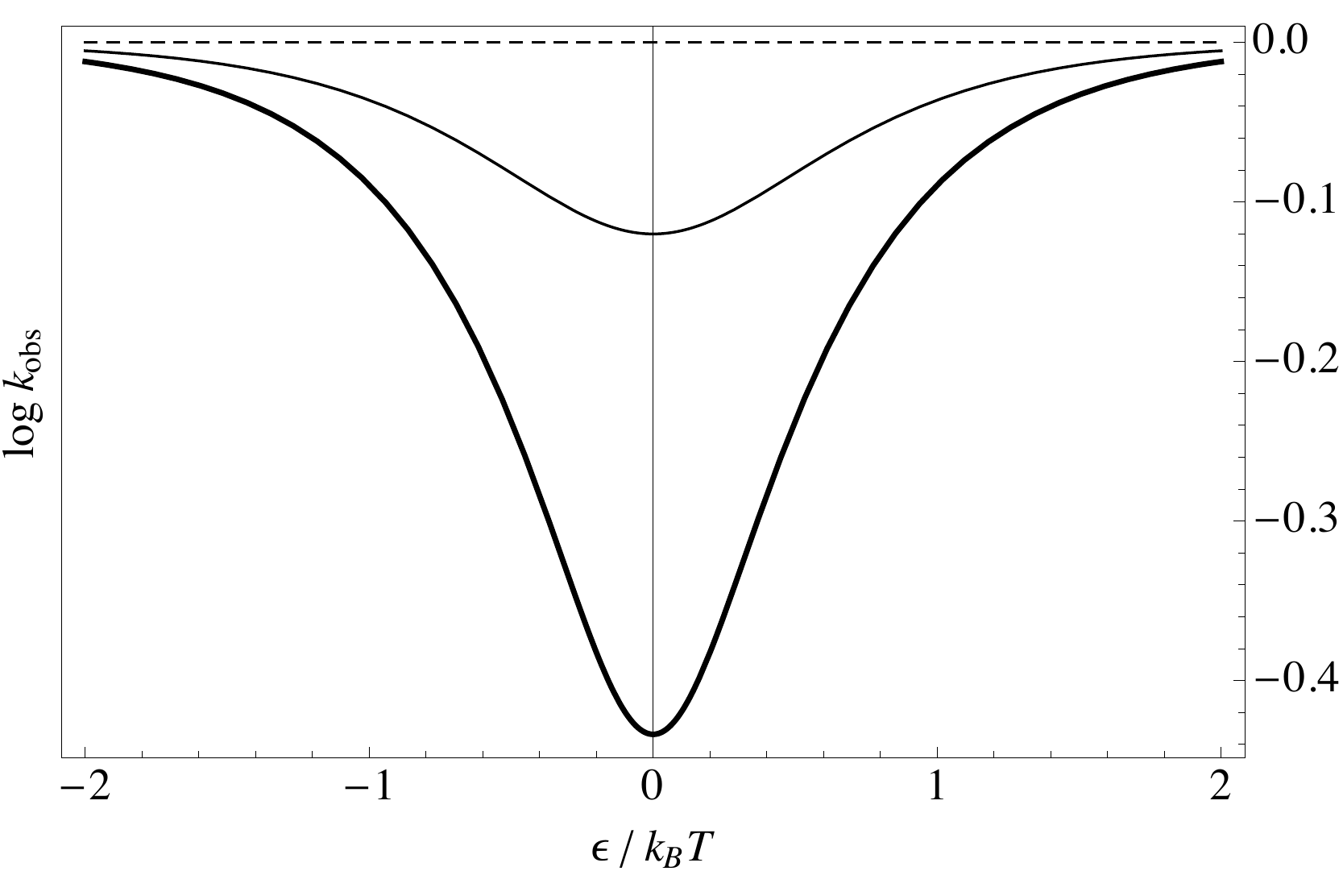}
\end{center}
\caption{
The logarithm of the apparent rate constant $k_{\text{obs}}$ is plotted against the 
local free energy change $\bar{\epsilon}=\epsilon/\kBT$
for three different values of $\bar{J}$: 0 (dashed line), 1/4 (thin solid line) and 1/2 (thick solid line).
}\label{f:chev01}
\end{figure}

Equation~\eqref{e:n010} is used to convert the expression in terms of $\avg{\sigma}$
into that in terms of $\bar{\epsilon}$.
The result is cumbersome so it is not shown here.
In Fig.~(\ref{f:chev01}), $\log k_{\text{obs}}$ is plotted against $\bar{\epsilon}$.
Three different values of $\bar{J}$--- $0$, $1/4$ and $1/2$ --- were used.
They are shown in the thin dashed line, the thin solid line, and the thick
solid line, respectively.
Notice that in the present model positive $\epsilon$ favors the native state,
so the plot is reflected with respect to the $y$-axis  compared to
the chevron plots with denaturant concentration as the $x$-axis.
This does not cause much trouble in reading this figure because it is symmetric
with respect to this reflection.

In the non-cooperative $\left(J=0\right)$ case, $k_{\text{obs}}$ is a constant.
As the cooperativity increases to $\bar{J}=1/2$, the chevron plot turns into a  V-shaped curve.
Notice that both arms of the chevron plot are not straight lines.
It is similar to most of the rolled-over chevron plots%
\cite{bc30:10428,proteins30:2,prl90:258104,pnas107:2920}.
In this constant NNI Ising model, two important observations can be made.
First, without cooperativity, this system does not have a V-shape chevron plot.
Conversely, with certain degree of cooperativity, the V-shape feature is intrinsic.
Second, in the figure, it can be seen that when $\epsilon/\kBT\approx0$,
the chevron plot depends strongly on the cooperativity $J$.
When the external conditions do not favor either the native or the denatured states, 
the kinetics is dominated by the cooperative couplings.
From this view point, near the bottom of the V-shape curve,
it is not the two-state like behavior of the system which results in the shape of the curve.
In contrast, the far ends of the two rolled-over wings represents the
behavior of a set of almost independent two-level systems.
Obviously, many real or model systems do not show chevron plot rollover due to
the same reason discussed here%
\cite{bc30:10428,proteins30:2,prl90:258104,pnas107:2920}.
Besides, the two wings of the chevron plot usually roll over with quite different trends
so the plot is often asymmetric.
This is totally within expectation because the energetics and couplings in a real
system is much more complicated than a uniform Ising model.
A protein system contains many different domains,
for example helix structures and sheet structures, with different
folding-unfolding transition energetics and cooperative couplings.
Even for small peptides with only one major secondary structure,
it can be understood that the cooperative coupling between two denatured SUs
may be quite different from that between two native SUs.
In addition, if the abscissa is one of the external conditions such as denaturant concentration,
instead of free energy, the shape of the curve can also change if the free energy
does not depend on the external condition linearly.
These variations are already as simple as possible, but they are not included in
the present model.
Nevertheless, if the degrees of freedom in such biological macromolecules
can be wisely grouped into several subsystems of rather uniform Ising-like systems,
each of the subsystems will have their own thermodynamic and kinetic features
similar to the model system discussed in the present work%
\cite{jmb343:223,pnas102:4741,ps16:449,bj94:4828}.
Thus, this model will be useful for understanding the composite systems.

Another important point is that, in order to reproduce qualitatively how the
kinetics and chevron plots look like, the cooperativity is usually weak.
The value of $\bar{J}=J/\kBT=1/2$ used in most of the plots is already rather high
that the non-exponential relaxation looks exaggerated.
However, it had long been considered that rather strong cooperative couplings exist
between units of proteins and of other biopolymers\cite{apc23:121}.
Some systems even have very high cooperativity, as evident from thermodynamic
studies\cite{qrb35:111,qrb35:205}.
There is no doubt about the ubiquity of strong cooperativity in biopolymers.
However, strongly cooperative degrees of freedom may go through configuration transition
at very different time-scale than weakly cooperative degrees of freedom.
In our analysis, for example, if the characteristic transition rate of individual SUs is the same,
the stronger the positive cooperativity, the longer is the frustrating period during
global configuration relaxation.
Monitoring the kinetics at different time-scales, it may be the configuration transition of
either only the strongly cooperative part or only the weakly cooperative part of the system that
is observed.
Moreover, the system can be quantitatively separated into different units.
For example, imagine two long $\beta$-strands that are bound together in the
native structure (a long hairpin with the turn ignored).
If one hydrogen-bonding-pair candidate already form hydrogen bond,
the ``nearest neighboring'' h-bonding-pair candidates are already so close
to their native structure that the chance of the cooperative formation of these 
neighboring h-bonds is really big.
This cooperative coupling also extends over at least several neighboring h-bonding candidates.
As a result, the `flipping' of any single h-bonding candidate is difficult.
However, if the system can be approximately diagonalized with the strongest cooperative coupling
included, a set of delocalized structure units will emerge, and the residual couplings
between these new units become weak.
In the $\beta$-strand pair example above, the delocalized configuration units may be
similar to molecular excitons in aggregates.
This picture will not conflict with the known facts of the existence of strong cooperative coupling.
Meanwhile, abstract one-dimensional Ising-like model will be quite sufficient for exploring and 
revealing the thermodynamic and kinetic characteristics of the delocalized structure units.

\section{Conclusions}
The Ising model is valuable for qualitative and conceptual investigations of many
complicated statistical physical systems.
Protein folding problem, among others, had benefited from the development of
Ising-like models aiming at incorporating the crucial features of the biopolymers.
In most of the Ising-like models, exact solutions, even just the equilibrium ones,
are difficult to obtain\cite{prl88:258101},
not to mention kinetics.
In this work, the focus was put on the one-dimensional Ising model with
uniform local free energy change and constant nearest-neighbor interaction.
Some of the relevant results in early literature treating this model with
the quasi-chemical (QC) approximation were collected.
Both equilibrium and kinetics were well studied.
Those results showed that the steady-state solution of the kinetic QC kinetic Ising model
(QCKI) is also the correct exact equilibrium solution.
Along this line, the contribution of this work was to extend the QC approximation
and found reasonable definitions of  the activities and activities coefficients
that carry all of the information about the cooperativity (and therefore nonideality).
In order to better understand the meaning of the kinetics of the model system,
it was found that introduction of a novel quantity called instantaneous activity
and the corresponding instantaneous activity is extremely useful.
Especially, the kinetics can be rearranged into physically meaningful forms.
The deviation of the unfolding/refolding relaxation process from simple exponential
can be clearly analyzed.
If it is assumed that initial-rate method can be seriously applied to monitor the
rate of the unfolding/refolding reactions, this model can also predict the form of the
chevron plots.
In this model, chevron rollover is an essential feature.
The remaining questions are how to relate the local free energy change with
external physical conditions, and whether the energetics and couplings in the system
are uniform enough to correspond better to the model.

Some other important thermodynamic and kinetic features of this model had not been
covered in this work, due to the subjective preference of the author itself.
For example, temperature-dependent folding/unfolding processes are of great value.
To investigate those kinetics, it will be nice if the calorimetric experiments can also be
simulated to confirm the physical picture\cite{jbp38:543}.
The present model, although abstract and simple, remains to be one of the
most interesting model of protein folding.

\section*{Acknowledgment}
The author wishes to thank Prof.~Sheng~Hsien~Lin for many useful discussions.

\bibliographystyle{apsrev4-1}
\bibliography{quasichem}
\end{document}